\documentclass{iau}

\usepackage{amsmath}
\usepackage{graphicx}
\usepackage{multirow}

\begin{document}

\lefttitle{A. A. Soemitro et al.}
\righttitle{Precision spectrophotometry for PNLF distances: the case of NGC 300}

\journaltitle{Planetary Nebulae: a Universal Toolbox in the Era of Precision Astrophysics}
\jnlDoiYr{2023}
\doival{10.1017/xxxxx}
\volno{384}

\aopheadtitle{Proceedings IAU Symposium}
\editors{O. De Marco, A. Zijlstra, R. Szczerba, eds.}
 
\title{Precision spectrophotometry for PNLF distances: \\
the case of NGC 300}

\author{Azlizan A. Soemitro$^{1,2}$, Martin M. Roth$^{1,2}$, Peter M. Weilbacher$^{1}$, Robin Ciardullo$^{3}$, George H. Jacoby$^{4}$, Ana Monreal-Ibero$^{5}$, Norberto Castro$^{1}$, Genoveva Micheva$^{1}$}
\affiliation{
$^{1}$Leibniz-Institut für Astrophysik Potsdam (AIP), Germany \\
$^{2}$Universität Potsdam, Germany \\
$^{3}$Department of Astronomy \& Astrophysics, The Pennsylvania State University, USA\\
$^{4}$NSF’s NOIRLab, Tucson, USA\\
$^{5}$Leiden Observatory, Leiden University, The Netherlands \\
}

\begin{abstract}
The Multi-Unit Spectroscopic Explorer (MUSE) has enabled a renaissance of the planetary nebula luminosity function (PNLF) as a standard candle. In the case of NGC 300, we learned that the precise spectrophotometry of MUSE was crucial to obtain an accurate PNLF distance. We present the advantage of the integral field spectrograph compared to the slit spectrograph in delivering precise spectrophotometry by simulating a slit observation on integral field spectroscopy data. We also discuss the possible systematic shift in measuring the PNLF distance using the least-square method, especially when the PNLF cutoff is affected by small number statistics.

\end{abstract}


\maketitle

\section{Introduction}

The planetary nebula luminosity function (PNLF) is a secondary distance indicator with $\sim$10\% accuracy \citep{2010PASA...27..149C, 2022FrASS...9.6326C}. Using narrowband photometry, one can obtain the V-band equivalent [OIII]$\lambda$5007 magnitude defined by \citet{1989ApJ...339...39J}:

\begin{equation}
    m_{5007} = -2.5 \: \mathrm{log} \: F_{5007} - 13.74
\end{equation}

\noindent
The number distribution of planetary nebulae (PNe) per magnitude bin for a given galaxy can then be modelled using the empirical law described by \citet{1989ApJ...339...53C}:

\begin{equation}
    N(M) \propto e^{0.307M}\{1-e^{3(M^*-M)}\}
\end{equation}

\noindent
Variations of the formula have been introduced to explain the influence of different underlying stellar populations \citep{2013A&A...558A..42L, 2015A&A...575A...1R}. However, these formulae merely affect the faint end of the PNLF. The definition of the PNLF cutoff, which is crucial for the distance determination, remains unchanged. 

The renaissance of the PNLF as standard candles came with the use of the Multi-Unit Spectroscopic Explorer \citep[MUSE;][]{2010SPIE.7735E..08B} on the Very Large Telescope (VLT). As an integral field spectrograph (IFS), it allows simultaneous imaging and spectroscopy to find and classify PNe, even in more crowded regions of galaxies, that previously were not accessible using the classical narrow-band imaging methods \citep{2017ApJ...834..174K, 2020A&A...637A..62S, 2021arXiv210501982R, 2022MNRAS.511.6087S, 2023A&A...671A.142S}.

The precision of the $m_{5007}$ defines the precision of the PNLF cutoff of a galaxy, which consequently defines the accuracy of the distance determination. To achieve this, \citet{2021arXiv210501982R} developed the differential emission line filter (DELF), which suppresses systematic errors and delivers higher signal-to-noise ratios for the photometry, with a typical photometric error of $\sim$ 0.04 mag. They demonstrated that their method was able to reach distances up to $\sim$ 40 Mpc, twice as far as most studies done with the classical narrow-band photometry until the early 2010s. 

\section{Planetary nebulae in NGC 300}

NGC 300 is a nearby spiral galaxy in the foreground of the Sculptor Group, where a few PN surveys have already been conducted, making it a good case to test the DELF technique using MUSE data \citep{2023A&A...671A.142S}. Previously, using narrow-band imaging on the New Technology Telescope (NTT), \citet{1996A&A...306....9S} found 34 PNe, covering $3 \times 2.2 \times 2.2$ arcmin$^2$. Another survey with the FORS2 instrument on the VLT found a total of 104 PNe using the narrow-band imaging mode within $2 \times 6.8 \times 6.8$ arcmin$^2$ \citep{2012A&A...547A..78P}. For the brighter 32 PN candidates, they obtained spectroscopic follow-up data using the multi object spectroscopy mode of the same instrument (MXU-FORS2). Using the MUSE instrument, \citet{2018A&A...618A...3R} observed 7 fields in the wide field mode (WFM, $1 \times 1$ arcmin$^2$ FoV) that cover the nucleus, parts of a spiral arm, and parts of a nearby inter-arm region, as a pilot study for crowded field spectroscopy, which explored a plethora of possible science cases. They identified 45 PNe, which was not ideal to construct a proper PNLF. As a follow up for the PN science case, \citet{2023A&A...671A.142S} employed additional archival MUSE data and increased the dataset from 7 fields to a total of 44 fields, and identified 109 PNe. Using the DELF, the study presented a similar number of PNe in comparison to the previous study by \citet{2012A&A...547A..78P}, however needing only half of the surveyed area.

\section{Precision spectrophotometry: integral field spectrograph vs. slit spectrograph}

For the common PNe in the sample, \citet{2023A&A...671A.142S} have compared the $m_{5007}$ measurements with the results from \citet{1996A&A...306....9S} and \citet{2012A&A...547A..78P}. While their $m_{5007}$ values were consistent with the earlier study done with the narrow-band imaging photometry on the NTT, they found that the results from the FORS2 survey study were fainter by an average of 0.71 mag, a discrepancy factor of $\sim$ 2. The $m_{5007}$ comparison can be seen in Figure 1. The speculation was, since \citet{2012A&A...547A..78P} calibrated their photometry using spectroscopic fluxes obtained with the multi object spectrograph, that they might be subject to slit loss effects \citep{2023A&A...671A.142S}. Due to atmospheric dispersion, one has to orient the slit along the parallactic angle to obtain precise spectrophotometry \citep{1982PASP...94..715F, 1993ApJ...417..209J}. This is necessary to minimise the slit loss, because the wavelength dependence of the point spread function full width half maximum (PSF FWHM) might complicate the slit positioning. For instance, if the slit was positioned based on the image in the R-band, the slit position could be off-centre at [OIII]$\lambda$5007. Moreover, the PSF FWHM also increases with shorter wavelengths \citep{1966JOSA...56.1372F, 1978JAtS...35.2236B, 2013A&A...549A..71K}, which might causes more loss of flux in blue.

  \begin{figure}
   \centering
   \includegraphics[width=0.65\hsize]{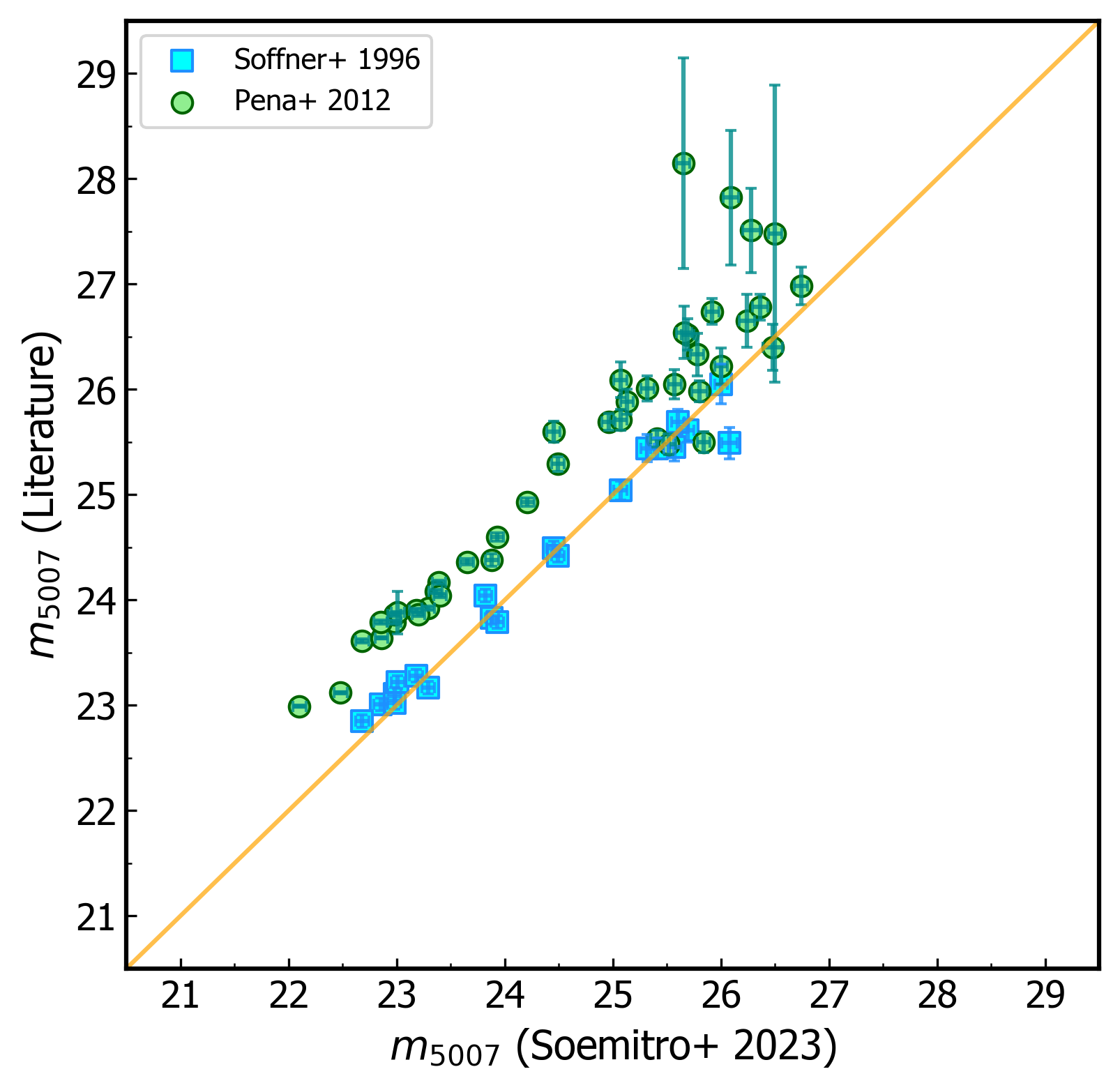}
      \caption{Comparison of the $m_{5007}$ for different PN surveys in NGC 300. The magnitudes from \protect\citet{2012A&A...547A..78P} is 0.71 mag fainter than the other surveys. The plot is adapted from \protect\citet{2023A&A...671A.142S}.}
         \label{fig:fields}
   \end{figure} 

To demonstrate the effect of atmospheric dispersion, we used a standard star observation  obtained with another integral field spectrograph, the Potsdam Multi-Aperture Spectrophotometer \citep[PMAS;][]{2005PASP..117..620R} whose raw data comes without atmospheric dispersion correction. The star was observed in the lens array mode with 1"/pixel sampling (LARR, $16 \times 16$ arcmin$^2$ FoV) and a wavelength coverage of $\sim 3500 - 6800$ Å (V600 Grating, GROT=144.5). We took two wavelength slices: 3800 Å and 6500 Å. We created a rectangular aperture to simulate a slit, using the width of two pixels. The simulated slit position was fixed based on the PSF centroid in 6500 Å, then we performed photometry for both slices. As a comparison, we also did photometry using circular apertures with the radius of two pixels, however, we adjusted their positions based on the PSF centroid in each respective data cube layer. Figure 2 displays the star with the apertures in the two wavelengths. In this configuration, we found that the slit loss in 6500 Å was $\sim$ 10\%. However, with $\sim$ 34\%, the light loss in the blue was much more severe.

  \begin{figure*}
   \centering
   \includegraphics[width=0.75\hsize]{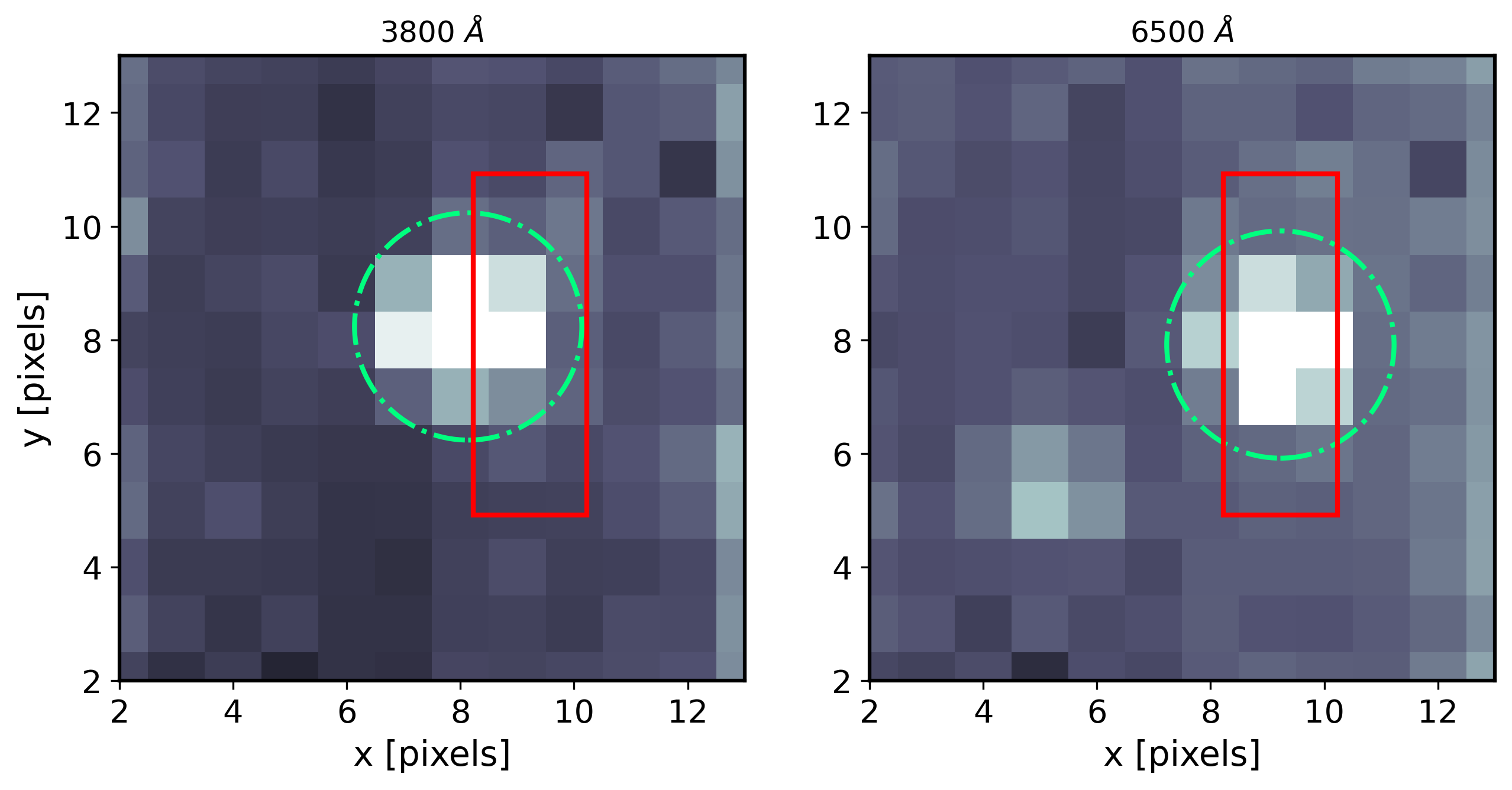}
      \caption{Simulated apertures for testing the light loss effect of atmospheric dispersion for a standard star. The dot-dashed green line represents measurements using IFS. The solid red line represent measurements using a slit spectrograph. This effect can be minimised by aligning the slit with the parallactic angle continuously during observations \protect\citep{1982PASP...94..715F, 1993ApJ...417..209J}.}
         \label{fig:fields}
   \end{figure*} 

While we cannot prove that the effect was exactly the cause of the discrepancies between \citet{2012A&A...547A..78P} and \citet{2023A&A...671A.142S}, since it was unclear whether the slit alignment with the parallactic angle was done or not for the FORS2 observations, we can plausibly argue that the effect can have large effects on the flux calibration, in particular knowing that the seeing FWHM can be quite variable over the course of a night. Moreover, we demonstrated that the IFS has an advantage to handle such atmospheric effects, delivering a more precise spectrophotometry and consequently better distances. In the case of MUSE data, the atmospheric dispersion as a function of wavelength is even corrected by lateral shifts that are applied through the data reduction pipeline \citep{2020A&A...641A..28W}. 

\section{The PNLF of NGC 300}

In \citet{2023A&A...671A.142S}, the PNLF distance to NGC 300 was derived using the maximum likelihood estimation method \citep{1989ApJ...339...53C}. The PNLF maximum likelihood distance estimates for both the MUSE and the FORS2 datasets resulted in $(m-M)_0 = 26.48^{+0.11}_{-0.26}$ and $(m-M)_0 =  27.30^{+0.09}_{-0.20}$, respectively, reflecting the photometric discrepancy explained above. However, in the original study of \citet{2012A&A...547A..78P}, they measured a distance of $(m-M)_0 = 26.29^{+0.12}_{-0.22}$, derived on the basis of the least-square fitting method. By using such a method, the distance measurement became very dependent on the binning. Since the PNLF  bins of NGC 300 presented only a small number of PNe, the exact definition of the bins has likely caused a systematic shift to the distance measurement \citep{2023A&A...671A.142S}. The choice of binning also smeared out the details of the PNLF shape. In their study, \citet{2012A&A...547A..78P} reported that they did not observe any PNLF dip, that however is typically being observed in star forming galaxies \citep{2002AJ....123..269J}. With a higher resolution of the magnitude bins, \citet{2023A&A...671A.142S} showed that the dip can be seen even in the FORS2 data, and also in the MUSE observations, as shown in Figure 3. With the precise spectrophotometry delivered by MUSE, the PNLF distance was brought into good agreement with the Cepheid and the tip of the red giant branch (TRGB) distances. The distance comparison is shown in Figure 4.

  \begin{figure}
   \centering
   \includegraphics[width=0.6\hsize]{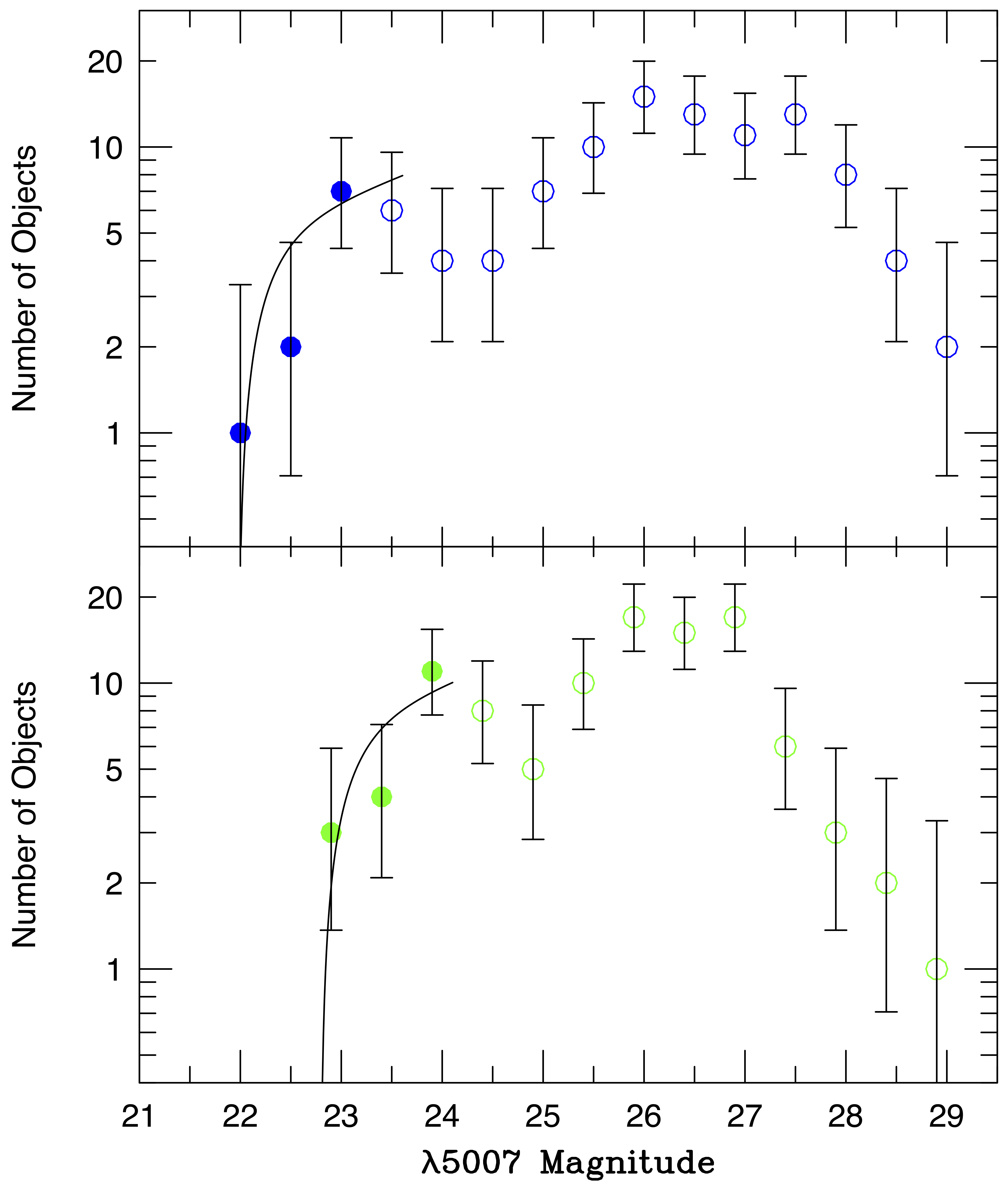}
      \caption{The PNLF for MUSE data \protect\citep{2023A&A...671A.142S} in blue and FORS2 data \protect\citep{2012A&A...547A..78P} in green. The plot is adapted from \protect\citet{2023A&A...671A.142S}.}
         \label{fig:fields}
   \end{figure} 

  \begin{figure}
   \centering
   \includegraphics[width=0.8\hsize]{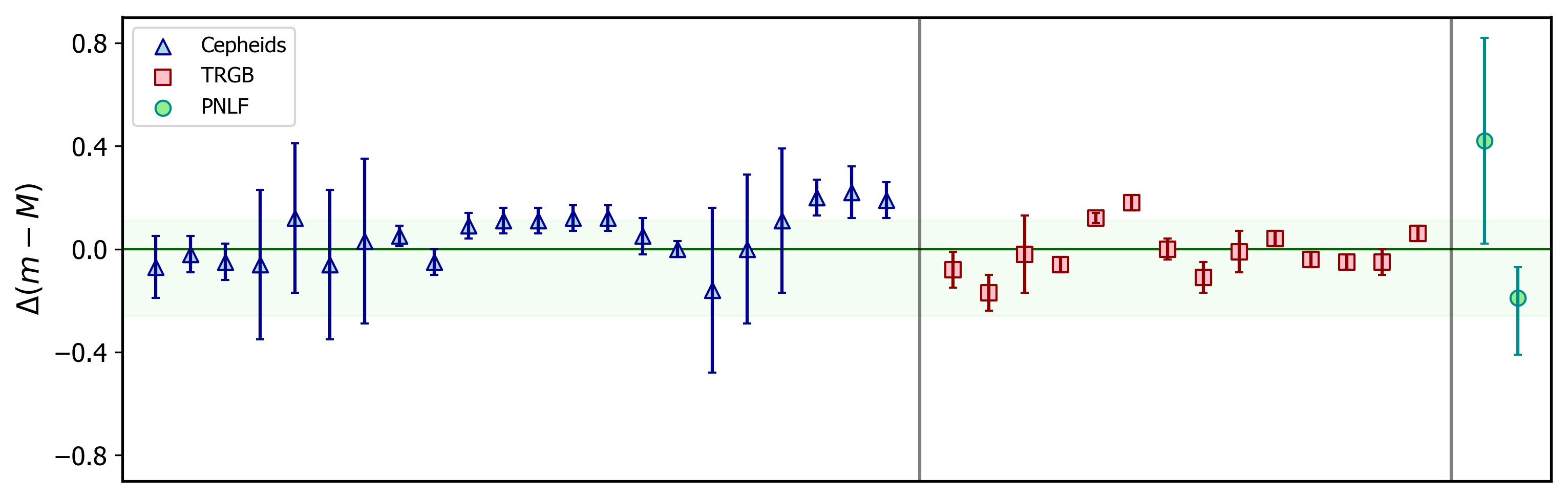}
      \caption{The MUSE-PNLF distance and its uncertainty is presented as a solid green line and the green shaded range, respectively. As a comparison with distances from literature, Cepheids (blue triangles), tip of the red giant branch (red squares), and previous PNLF distance estimates (the two green circles) are shown. A full reference list of literature distances can found in \protect\citet{2023A&A...671A.142S}.}
         \label{fig:fields}
   \end{figure}

\section{Summary}
The case of NGC 300 has shown that the precision of spectrophotometry can have a significant impact on the distance measurement. We made a test using another integral field spectrograph, PMAS, to show the probable effect of atmospheric dispersion when using a slit spectrograph to do the spectrophotometry. This illustration supports the importance of using an IFS for measuring accurate $m_{5007}$ magnitudes for PNLF distance determinations. Moreover, the methodology of how to fit an analytical PNLF to the observed photometry is crucial. In case of small number statistics at the PNLF cutoff, using the least-square minimisation method that is dependent on the bin size, is likely to bias the distance measurement. Therefore, the maximum likelihood approach is recommended. As the only IFS that to date has the sensitivity and spatial resolution to detect PNe in distant galaxies, the MUSE instrument is crucial for the measurement of PNLF distances in the future. With the currently demonstrated range of up to $\sim$ 40 Mpc, the PNLF can be assumed to be an additional calibration tool for SN Ia distances and a contribution to the ongoing discussion of the Hubble tension \citep{2023arXiv230911603J}.

\bibliographystyle{aa}
\bibliography{pnlf}

\end{document}